# Pumping Heat with Quantum Ratchets

T. E. Humphrey[a], H. Linke[a,b], R. Newbury[a]

[a] School of Physics, University of New South Wales, UNSW Sydney 2052, Australia

[b] Physics Department, University of Oregon, Eugene OR 97403-1274 USA

Email: hl@phys.unsw.edu.au, Ph: +61 2 9385 5928, Fax: +61 2 9385 6060

PACS codes: 05.60.+w, 73.23.Ad, 73.50.Fq , 73.50.Lw

Keywords: Non-linear, mesoscopic, quantum ratchet, heat pump

**Abstract**

We describe how adiabatically rocked quantum electron ratchets can act as heat pumps. In general, ratchets may be described as non-equilibrium systems in which directed particle motion is generated using spatial or temporal asymmetry. In a rocked ratchet, which may also be described as a non-linear rectifier, an asymmetric potential is tilted symmetrically and periodically. The potential deforms differently during each half-cycle, producing a net current of particles when averaged over a full period of rocking. Recently it was found that in the quantum regime, where tunnelling contributes to transport, the net current may change sign with temperature. Here we show that a Landauer model of an experimental tunnelling ratchet [Linke et. al., *Science* **286,** 2314 (1999)] predicts the existence of a net heat current even when the net particle current goes through zero. We quantify this heat current and define a coefficient of performance for the ratchet as a heat pump, finding that more heat is deposited in each of the two electron reservoirs due to the process of rocking than is pumped from one reservoir to the other by the ratchet.

## 1 Introduction

In general, a ratchet may be described as a non-equilibrium system in which directed particle motion is generated through the use of asymmetry [1-3]. Often, the non-equilibrium condition is achieved by varying an asymmetric potential with time. In this case the asymmetry defines a preferred direction, while the time variation provides the source of energy necessary to create a net current of particles [3]. One such example is a 'rocked' ratchet, in which an asymmetric potential is tilted symmetrically and periodically. The potential deforms differently during each half cycle of tilting, so that a net current is produced when the current is averaged over a full period of rocking [4]. Throughout the present paper we will only discuss so-called "adiabatically" rocked ratchets [5], in which symmetric tilting occurs on time scales much slower than all other time scales of the system. Such a system is essentially a non-linear rectifier. In the classical regime a net current can only occur when the *height* of the potential barrier depends upon the tilt direction. In the quantum regime however, where wave reflection and tunnelling can occur, not only the height, but also the *shape* of the barrier becomes important. In particular, narrow barriers will transmit more tunnelling particles than wide barriers, and smooth barriers, which cause less wave reflection than sharp barriers, will allow more high energy particles to pass over the barriers. It then follows that a change in the shape of the asymmetric barrier in a quantum ratchet as a result of tilting is sufficient to produce a net current, even when the barrier height remains constant. This quantum net current has been found to change direction with temperature [5,6]. The origin of the temperature dependent behaviour is illustrated in Fig. 1, which shows an asymmetric energy barrier between two-dimensional electron reservoirs. A square-wave voltage of amplitude $V_0$ is applied to 'rock' the potential. When tilted in one direction, the asymmetric barrier deforms to be thicker and smoother (Fig. 1a), suppressing tunnelling, but also reducing wave reflection of electrons, so favouring transmission of electrons with high energies. When it is tilted in the other direction however, the potential deforms to be thinner and sharper (Fig. 1b), enhancing tunnelling but increasing wave reflection, thus favouring the transmission of low energy electrons. In this way, the two contributions to the net current, tunnelling through and excitation over the energy barrier, flow, on average, in opposite directions. By tuning the temperature, rocking voltage or Fermi energy such that one of these two contributions exceeds the other, the net current direction can be chosen. In the present paper we point out that at parameter values where the contributions of the two components of the net *particle* current are equal and opposite (that is, where the net particle current goes through zero), a net *energy* current still exists because the average energy transported in each direction is not the same. In the following we briefly describe the experimental results of [6] and introduce a Landauer model for this experiment which will allow us to quantify the heat current generated by the ratchet.

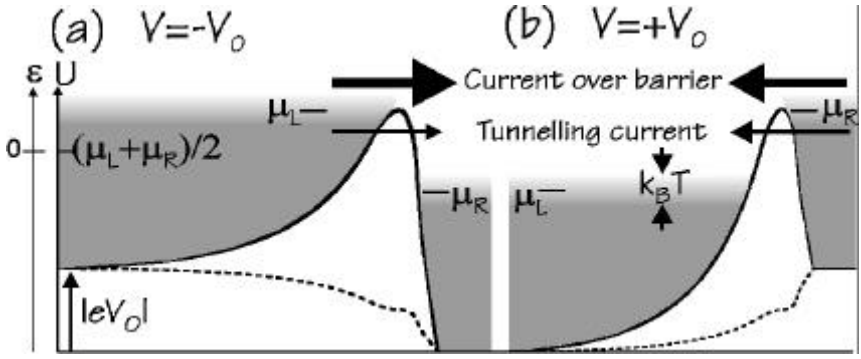

**Fig 1:** A rocked electron ratchet. The solid lines are an estimation of the confinement potential experienced by electrons as they traverse the experimental ratchet wave-guide shown in Fig 2 (inset). The Fermi distribution of electrons as a function of energy is indicated by the grey regions, where lighter grey corresponds to an occupation probability of less than one. The boldness of the arrows is indicative of the relative strengths of the contributions of high and low energy electrons to the current across the barrier under negative (a) and positive (b) voltages. The dashed lines indicate the spatial distribution of the assumed voltage drop over the barrier, which is scaled with the local potential gradient of the barrier at zero voltage.

**2 Experimental quantum ratchet**

A scanning electron microscope (SEM) image of the experimental quantum ratchet device is shown in Fig. 2 (inset). The darker areas are trenches which were defined by shallow wet etching and electron-beam lithography in a two-dimensional electron gas (2DEG) AlGaAs/GaAs heterostructure. This process created an asymmetric one-dimensional (1D) wave-guide connecting 2D electron reservoirs. The crucial feature of the ratchet is the asymmetric point contact on the right, which can be adjusted in width by applying a voltage to the 2DEG areas above and below the right point contact that serve as side gates (marked SG in the SEM image). The side gate voltage tunes the energy of the 1D wave modes, effectively creating an asymmetric energy barrier which is experienced by the electrons as they traverse the wave-guide. The left point contact, which is not influenced by the side gates, plays no significant role in determining the behaviour of the device as a ratchet. The dimensions of the device (~1 μm) were much smaller than the length scales for elastic (6 μm) and inelastic (>10 μm) scattering at the temperatures and voltages used in the experiment ($k_BT$ and $eV_0 \leq 1$ meV).

A low-frequency square-wave voltage of amplitude $V_0$ was applied between the two electron reservoirs to adiabatically rock the device, and the resulting net current, averaged over many periods of rocking, was measured using phase locking techniques. The direction of the net current was found to depend upon temperature, rocking amplitude, and the applied gate voltage. In Fig. 2 we show measurements of the net current versus the amplitude of the rocking voltage for various temperatures at constant side-gate voltage. For small voltages all three curves display parabolic behaviour, as expected for a lowest order non-linear effect, which at the lowest two temperatures (0.6 K and 2 K), turns over to reverse direction at a rocking voltage amplitude of $V_0 \approx 1$ mV. These results may be interpreted by referring to Fig. 1, which illustrates the idea that high energy electrons and tunnelling electrons travel, on average, in opposite directions. At low rocking voltage and temperature a positive net electrical current is measured (corresponding to a current of electrons from right to left in Fig. 1), indicating that tunnelling electrons dominate the net current. As either the temperature or voltage is increased, the energy range of electrons which contribute to transport widens, resulting in a greater contribution from electrons with energies higher than the barrier, leading to a negative net current. When the two contributions are equal in magnitude, the net current undergoes a sign reversal.

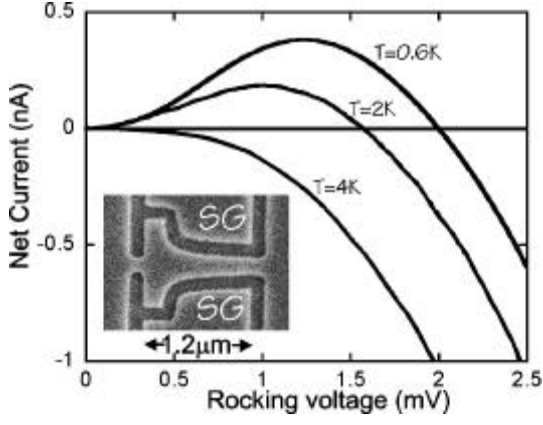

**Fig. 2:** Main: Measured net current as a function of rocking amplitude at a number of temperatures as indicated. Reversals in the direction of the net current as a function of rocking amplitude, and implicitly as a function of temperature, are observed. Data taken from [6]. **Inset:** A scanning electron microscope image of the ratchet device (top view). The dark regions are etched trenches that electrically deplete a two-dimensional electron gas located at the AlGaAs/GaAs interface beneath the surface, forming a one-dimensional wave-guide. Due to quantum confinement inside the waveguide, an electron moving from left to right will experience an asymmetric potential barrier similar to that shown in Fig. 1. Note the side gates (marked SG) which are used to tune the height of the potential barrier which is experienced by electrons moving though the ratchet. The left point contact does not play a significant role in the behaviour of the device as a ratchet.

## 3 The Landauer model

The Landauer equation expresses the current flowing through a mesoscopic device between two reservoirs as a function of the Fermi distribution of electrons in the reservoirs and of the energy dependent probability that an electron will be transmitted through the device [7]. It may be written as:

$$I = \frac{2e}{h} \int_{-\mu_{av}}^{\infty} t(\varepsilon,V)[f_R(\varepsilon,V) - f_L(\varepsilon,V)] d\varepsilon \,, \qquad (1)$$

where

$$f_{L/R}(\varepsilon,V) = \frac{1}{1 + \exp\left(\frac{-\varepsilon \pm eV/2}{k_B T}\right)} \qquad (2)$$

are the Fermi distributions in the left and right reservoirs (the upper/lower symbol in $\pm$ in all equations corresponds to the left (L) and right (R) reservoirs, respectively). $\varepsilon$ is the energy of the electrons, for convenience chosen relative to the average of the chemical potentials on the left and right sides, $\mu_{av} = (\mu_L + \mu_R)/2$ (Fig. 1). $T$ is the temperature of the 2DEG, $V$ is the applied bias voltage and $e = +1.6\,10^{-19}$C. Lastly, $t(\varepsilon,V)$ is the probability that electrons are transmitted across the barrier at a given bias voltage.

Eqn. (1) assumes that no inelastic scattering occurs inside the device. In addition, we require the applied bias to be much smaller than the Fermi energy. This means that the difference between the Fermi distributions will be negligible at low energies, and allows us to use $-\mu_{av} = -0.5(\mu_L + \mu_R)$ as the lower limit of integration, independent of the voltage sign. Non-linear effects, which form the basis of this particular ratchet effect, have been taken into account by solving the 1D Schrödinger equation to find $t(\varepsilon,V)$ for each positive and negative bias voltage individually. In order to do this, the energy of the lowest mode of the experimental wave guide (Fig. 2, inset) was estimated, resulting in the energy barrier shown in Fig. 1 (for more details see [6]). The height of the barrier corresponds to the confinement energy for lowest mode electrons at the narrowest point in the constriction.

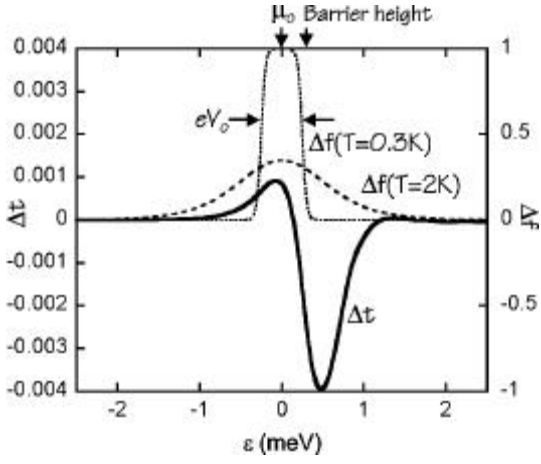

**Fig. 3:** The bold curve (corresponding to the left vertical axis) is the difference between the transmission probabilities for +0.5mV and –0.5mV tilting voltages as a function of electron energy. The dotted and dashed curves (corresponding to the right vertical axis) are the Fermi 'windows', $\Delta f(\varepsilon, V_0 = 0.5\text{ mV})$, centred on an equilibrium Fermi energy, $\mu_0 = 11.7$meV, for temperatures of 0.3K and 2K respectively. As the temperature is changed from 0.3K to 2K, note that the integral of the product of $\Delta f$ and $\Delta t$ will make a transition from positive to negative, leading to a reversal in the direction of the net particle current (Eq. 4). Small oscillations in $\Delta t$ exist for $\varepsilon > 1$ meV.

To obtain the barrier shape at finite voltage, an assumption about the spatial distribution of the voltage drop needs to be made. Arguing that a smooth potential variation can be approximated by a series of infinitesimally small steps, and that a step-like potential change may be assumed to cause a corresponding step-like voltage drop [8], we distribute the voltage drop in proportion to the local derivative of the barrier [9]. It is important to stress that the qualitative quantum behaviour of the ratchet does not depend upon the details of the voltage drop. The present choice, however, has the desirable side-effect that the barrier height remains independent of the sign of the voltage, resulting in the suppression of the classical contribution to the net current.

In the present model we include only contributions to transport from the lowest wave mode. The contribution from higher modes is qualitatively similar, and also negligibly small when the Fermi energy is approximately equal to the height of the barrier.

The net current is defined as the time average of the current over one period of rocking with a square-wave voltage of amplitude $V_0$:

$$I^{net} = \frac{1}{2}[I(V_0) + I(-V_0)]. \qquad (3)$$

This can also be written as:

$$I^{net} = \frac{e}{h} \int_{-\mu_{0v}}^{\infty} \Delta t(\varepsilon, V_0) \Delta f(\varepsilon, V_0) d\varepsilon, \qquad (4)$$

where $\Delta f(\varepsilon, V_0) \equiv f_R(\varepsilon, V_0) - f_L(\varepsilon, V_0)$, the 'Fermi window', is the difference between the Fermi distributions on the right and left of the barrier, and gives the range of electron energies which will contribute to the current. The Fermi window is centred on $\varepsilon = 0$ and has a width which depends upon the bias voltage and the temperature of the 2DEG (Fig 3). The term $\Delta t(\varepsilon, V_0) \equiv t(\varepsilon, V_0) - t(\varepsilon, -V_0)$, also shown in Fig. 3, is the difference between the transmission probabilities for an electron with energy $\varepsilon$ under positive (Fig. 1b) and negative (Fig. 1a) voltages. Electrons with energies under the barrier height are more likely to flow from right to left when the barrier becomes thinner (under positive voltage, Fig. 1, right), than from left to right when the barrier becomes thicker (under negative voltage, Fig. 1, left). This results in $\Delta t$ being negative in this energy range and then positive for energies above the barrier height where above situation is reversed. As $\Delta f$ is adjusted (through changing $T$, $V_0$ or Fermi energy, $\mu_0$) to sample the $\Delta t$ curve where it is negative rather than positive, the net current will change sign from positive to negative.

**4 Energy current**

The heat change associated with the transfer of one electron to a reservoir with chemical potential $\mu$ is given by [10]:

$$\Delta Q = \Delta U - \mu \qquad (5)$$

The internal energy, $\Delta U$, associated with the electron is taken with respect to the same global zero as the chemical potential, which is assumed to be unchanged by the electron transfer. The change in heat in the two reservoirs upon transfer of one electron from the right to the left is then given by $\Delta Q_{L/R} = \varepsilon + (\mu_L + \mu_R)/2 - \mu_{L/R} = \varepsilon \pm eV/2$. Note that the heat removed from one reservoir by an electron crossing the barrier, differs from the heat it adds to the other reservoir by $|eV|$, as a result of the kinetic energy acquired by the electron in the electric field driving the current.

The heat current entering the left and right reservoirs associated with the particle current generated by a voltage $V$ across the device is then obtained from the equation for the electrical current (Eq. 1). This is done by replacing the electron charge, $-e$, by a factor of $\Delta Q_{L/R}$ inside the integral. The heat current can then be written as

$$q_{L/R} = \mp \frac{2}{h} \int_{-\mu_{av}}^{\infty} (\varepsilon \pm eV/2) t(\varepsilon, V) \Delta f(\varepsilon, V) d\varepsilon \qquad (6)$$

The *net* heat current into the left and right reservoirs over a full cycle of square-wave rocking, $q_{L/R}^{net} = (1/2)[q_{L/R}(V_0) + q_{L/R}(-V_0)]$, is then:

$$q_{L/R}^{net} = \mp \frac{1}{h} \int_{-\mu_{av}}^{\infty} [(\varepsilon \pm eV_0/2) t(\varepsilon, +V_0) \Delta f(\varepsilon, +V_0) + (\varepsilon \mp eV_0/2) t(\varepsilon, -V_0) \Delta f(\varepsilon, -V_0)] d\varepsilon \qquad (7)$$

To obtain an intuitive understanding of the action of the ratchet as a heat pump at parameter values where the net particle current goes through zero, it is helpful to rewrite Eq. (7) as:

$$q_{L/R}^{net} = \mp \frac{1}{h} \int_{-\mu_{av}}^{\infty} \varepsilon \Delta t \Delta f d\varepsilon + \frac{1}{h} \frac{eV_0}{2} \int_{-\mu_{av}}^{\infty} t \Delta f d\varepsilon = \mp \frac{1}{2} \Delta E + \frac{1}{2} \Omega \qquad (8)$$

Here $t(\varepsilon, V_0) = (1/2)[t(\varepsilon, +V_0) + t(\varepsilon, -V_0)]$ is the average transmission probability for an electron under positive and negative bias voltage. $\Delta E = q_L^{net} - q_R^{net}$ is the heat pumped from the left to the right sides of the device due to the energy sorting properties of the ratchet, and can be non-zero only for asymmetric barriers. $\Omega = (V_0/2)[|I(+V_0)| + |I(-V_0)|]$ is the ohmic heating, averaged over one cycle of rocking. $\Delta E$ can be interpreted as the heat pumping power of the ratchet, averaged over a period of rocking, while $\Omega = q_L^{net} + q_R^{net}$ is the electrical power input, averaged over a period of rocking. We therefore define a coefficient of performance for the ratchet as a heat pump as:

$$c(T, V_0) = \frac{\Delta E}{\Omega} = \frac{q_R^{net} - q_L^{net}}{q_R^{net} + q_L^{net}} \qquad (9)$$

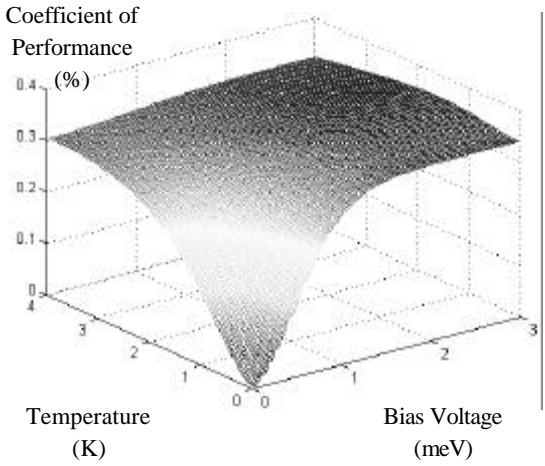

**Fig. 4:** The heat-pumping coefficient of performance of the ratchet model shown in Fig. 1, plotted as a function of rocking voltage and temperature. Each point on the surface corresponds to a set of values of rocking voltage, temperature and Fermi energy for which the net particle current goes through zero.

It is important to note that $\Delta E$ will be trivially non-zero when the net particle current is non-zero because each electron carries heat. This definition of $c$ therefore only makes sense for parameter values where the net particle current is zero.

To evaluate our model potential (Fig. 1) in terms of a heat pump, we have calculated $c$ for parameter sets where the net current goes through zero (Fig. 4). As required, the heat pumping power goes to zero for small bias and temperature, corresponding to the linear response limit where, by definition, the ratchet cannot work. The positive coefficient of performance indicates that heat is always pumped from left to right for the range of parameters used in the calculation. Small oscillations in $\Delta t$ at energies higher than the barrier exist and placing $\Delta f$ around these would result in heat being pumped from right to left. The fact that $c$ is small means that the total heat deposited in each reservoir due to ohmic heating is much larger than the heat pumped from the left to the right sides of the device. This means that, despite the heat pumping action of the ratchet, the internal energy of both reservoirs increases, but one reservoir is heated slightly less than the other. The experimental quantum ratchet may therefore be viewed as a poorly designed refrigerator, where half of the waste heat is deposited inside instead of outside the refrigerated region. For one reservoir to be cooled using the ratchet effect, $\Delta E$ would need to be larger than $W$ ($c > 1$), so that more heat was pumped out of one reservoir than was deposited there as a result of ohmic heating. The low coefficient of performance of the ratchet of [6] as a heat pump is a result of the fact that the potential barrier studied here transmits electrons with a wide range of energies in both rocking directions (all of which contribute to heating), while the ratio $\Delta t/t$ is less than 1%. Thus ohmic heating of each reservoir greatly exceeds the heat pumped from one side to the other. The heat pumping coefficient of performance of the ratchet would be enhanced by designing a potential which *only* transmitted electrons with energies higher than the equilibrium Fermi energy in one direction, and *only* transmitted electrons with energies lower than equilibrium Fermi energy in the other direction, so that $\Delta t/t \cong 1$. One way of achieving this may be to employ resonant tunnelling as a means of energy filtering. Resonant tunnelling barriers have in fact been predicted to be able to cool a reservoir when operated in DC mode [11]. An adaptation of this idea to rocked ratchets is currently under investigation.

This work was supported by the Australian Research Council.